\documentclass[10pt,a4paper,onecolumn]{article}
\usepackage{marginnote}
\usepackage{graphicx}
\usepackage{xcolor}
\usepackage{authblk,etoolbox}
\usepackage{titlesec}
\usepackage{calc}
\usepackage{tikz}
\usepackage{hyperref}
\hypersetup{colorlinks,breaklinks,
            urlcolor=[rgb]{0.0, 0.5, 1.0},
            linkcolor=[rgb]{0.0, 0.5, 1.0}}
\usepackage{caption}
\usepackage{tcolorbox}
\usepackage{amssymb,amsmath}
\usepackage{ifxetex,ifluatex}
\usepackage{seqsplit}
\usepackage{fixltx2e} 
\usepackage[
  backend=biber,
]{biblatex}
\bibliography{paper.bib}

\usepackage[top=3.5cm, bottom=3cm, right=1.5cm, left=1.0cm,
            headheight=2.2cm, reversemp, includemp, marginparwidth=4.5cm]{geometry}

\titleformat{\section}
  {\normalfont\sffamily\Large\bfseries}
  {}{0pt}{}
\titleformat{\subsection}
  {\normalfont\sffamily\large\bfseries}
  {}{0pt}{}
\titleformat{\subsubsection}
  {\normalfont\sffamily\bfseries}
  {}{0pt}{}
\titleformat*{\paragraph}
  {\sffamily\normalsize}

\usepackage{fancyhdr}
\makeatletter
\let\ps@plain\ps@fancy
\fancyheadoffset[L]{4.5cm}
\fancyfootoffset[L]{4.5cm}

\definecolor{linky}{rgb}{0.0, 0.5, 1.0}

\newtcolorbox{repobox}
   {colback=red, colframe=red!75!black,
     boxrule=0.5pt, arc=2pt, left=6pt, right=6pt, top=3pt, bottom=3pt}

\newcommand{\ExternalLink}{%
   \tikz[x=1.2ex, y=1.2ex, baseline=-0.05ex]{%
       \begin{scope}[x=1ex, y=1ex]
           \clip (-0.1,-0.1)
               --++ (-0, 1.2)
               --++ (0.6, 0)
               --++ (0, -0.6)
               --++ (0.6, 0)
               --++ (0, -1);
           \path[draw,
               line width = 0.5,
               rounded corners=0.5]
               (0,0) rectangle (1,1);
       \end{scope}
       \path[draw, line width = 0.5] (0.5, 0.5)
           -- (1, 1);
       \path[draw, line width = 0.5] (0.6, 1)
           -- (1, 1) -- (1, 0.6);
       }
   }

\patchcmd{\@maketitle}{center}{flushleft}{}{}
\patchcmd{\@maketitle}{center}{flushleft}{}{}
\patchcmd{\@maketitle}{\LARGE}{\LARGE\sffamily}{}{}
\def\maketitle{{%
  
  \AB@maketitle}}
\makeatletter
\renewcommand\AB@affilsepx{ \protect\Affilfont}
\renewcommand\AB@affilnote[1]{{\bfseries #1}\hspace{3pt}}
\makeatother

\renewcommand\Affilfont{\sffamily\small\mdseries}
\ifnum 0\ifxetex 1\fi\ifluatex 1\fi=0 
  \usepackage[T1]{fontenc}
  \usepackage[utf8]{inputenc}

\else 
  \ifxetex
    \usepackage{mathspec}
  \else
    \usepackage{fontspec}
  \fi
  \defaultfontfeatures{Ligatures=TeX,Scale=MatchLowercase}

\fi
\IfFileExists{upquote.sty}{\usepackage{upquote}}{}
\IfFileExists{microtype.sty}{%
\usepackage{microtype}
\UseMicrotypeSet[protrusion]{basicmath} 
}{}

\usepackage{hyperref}
\hypersetup{unicode=true,
            pdftitle={psrqpy: a python interface for querying the ATNF pulsar catalogue},
            pdfborder={0 0 0},
            breaklinks=true}
\usepackage{graphicx,grffile}
\makeatletter
\def\maxwidth{\ifdim\Gin@nat@width>\linewidth\linewidth\else\Gin@nat@width\fi}
\def\maxheight{\ifdim\Gin@nat@height>\textheight\textheight\else\Gin@nat@height\fi}
\makeatother
\setkeys{Gin}{width=\maxwidth,height=\maxheight,keepaspectratio}
\IfFileExists{parskip.sty}{%
\usepackage{parskip}
}{
\setlength{\parindent}{0pt}
\setlength{\parskip}{6pt plus 2pt minus 1pt}
}
\ifx\paragraph\undefined\else
\let\oldparagraph\paragraph
\renewcommand{\paragraph}[1]{\oldparagraph{#1}\mbox{}}
\fi
\ifx\subparagraph\undefined\else
\let\oldsubparagraph\subparagraph
\renewcommand{\subparagraph}[1]{\oldsubparagraph{#1}\mbox{}}
\fi

\title{psrqpy: a python interface for querying the ATNF pulsar catalogue}

        \author[1]{Matthew Pitkin}
    
      \affil[1]{Institute for Gravitational Research, SUPA, University of Glasgow,
University Avenue, Glasgow, UK, G12 8QQ}
  \date{\vspace{-5ex}}

\begin{document}
\maketitle

\marginpar{
  \sffamily\small

  {\bfseries DOI:} \href{https://doi.org/10.21105/joss.00538}{\color{linky}{10.21105/joss.00538}}

  \vspace{2mm}

  {\bfseries Software}
  \begin{itemize}
    \setlength\itemsep{0em}
    \item \href{https://github.com/openjournals/joss-reviews/issues/538}{\color{linky}{Review}} \ExternalLink
    \item \href{https://github.com/mattpitkin/psrqpy}{\color{linky}{Repository}} \ExternalLink
    \item \href{http://dx.doi.org/10.5281/zenodo.1175303}{\color{linky}{Archive}} \ExternalLink
  \end{itemize}

  \vspace{2mm}

  {\bfseries Submitted:} 16 January 2018\\
  {\bfseries Published:} 19 February 2018

  \vspace{2mm}
  {\bfseries Licence}\\
  Authors of papers retain copyright and release the work under a Creative Commons Attribution 4.0 International License (\href{http://creativecommons.org/licenses/by/4.0/}{\color{linky}{CC-BY}}).
}

\hypertarget{summary}{%
\section{Summary}\label{summary}}

This Python module provides an interface for querying the
\href{http://www.atnf.csiro.au/people/pulsar/psrcat/}{Australia
Telescope National Facility (ATNF) pulsar catalogue} (Manchester et al.
2005). The intended users are astronomers wanting to extract data from
the catalogue through a script rather than having to download and parse
text tables output using the standard web interface. It allows users to
access information, such as pulsar frequencies and sky locations, on all
pulsars in the catalogue. Querying of the catalogue can easily be
incorporated into Python scripts.

The module can also be used to create plots of pulsar period against
period derivative (\(P\) vs.~\(\dot{P}\) plots) using
\texttt{matplotlib} (Hunter 2007) as shown below.

\begin{figure}
\centering
\includegraphics{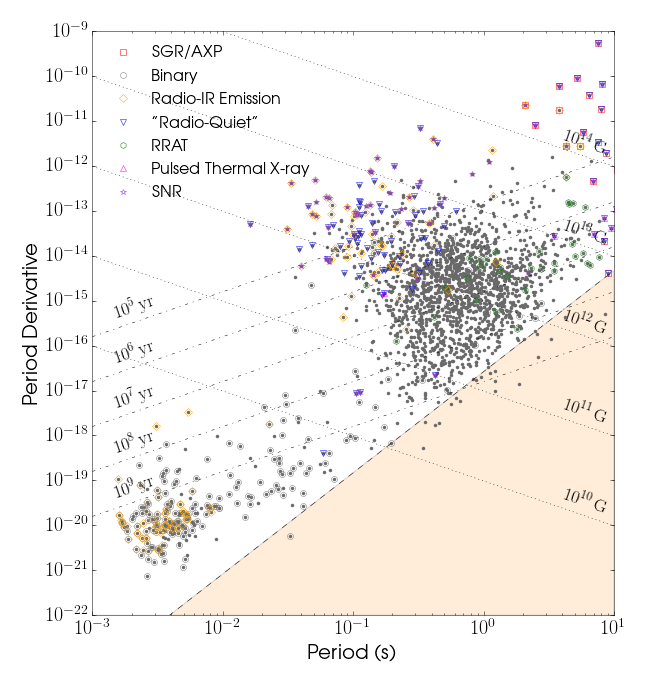}
\caption{A plot of pulsar period vs.~period derivative as produced using
\emph{psrqpy}}
\end{figure}

If requested the module can also return references for parameter values
for pulsars using the \texttt{ads} Python module (Sudilovsky et al.
2017).

Development of \emph{psrqpy} happens on Github (Pitkin 2017) and the
documentation is provided \href{http://psrqpy.readthedocs.io}{here}.

\hypertarget{refs}{}
\leavevmode\hypertarget{ref-matplotlib}{}%
Hunter, John D. 2007. ``Matplotlib: A 2D Graphics Environment.''
\emph{Computing in Science and Engineering} 9 (3): 90--95.
\url{https://doi.org/10.1109/MCSE.2007.55}.

\leavevmode\hypertarget{ref-ATNF}{}%
Manchester, R. N., G. B. Hobbs, A. Teoh, and M. Hobbs. 2005. ``The
Australia Telescope National Facility Pulsar Catalogue.''
\emph{Astronomical Journal} 129 (April): 1993--2006.
\url{https://doi.org/10.1086/428488}.

\leavevmode\hypertarget{ref-psrqpy_github}{}%
Pitkin, Matthew. 2017. ``Psrqpy on Github.'' 2017.
\url{https://github.com/mattpitkin/psrqpy}.

\leavevmode\hypertarget{ref-ADS}{}%
Sudilovsky, V., A. Casey, G. Barentsen, D. Foreman-Mackey, de Val-Borro.
M., and J. Elliott. 2017. ``The Ads Python Package.'' 2017.
\url{https://ads.readthedocs.io/}.

\end{document}